\documentclass[useAMS,usenatbib]{mn2e}

\usepackage{graphics}

\title[Contact binary stars]
{Masses and angular momenta of contact binary stars}

\author[K. D. Gazeas and P. G. Niarchos]
{K. D. Gazeas$$\thanks{e-mail: kgaze@physics.auth.gr; pniarcho@phys.uoa.gr} and
P. G. Niarchos$$\footnotemark[1]\\
$$Department of Astrophysics, Astronomy and Mechanics, Faculty of Physics, University of Athens, 
GR-157 84, Zografos, Athens, Greece\\
}

\date{Accepted 2006 April 13.
      Received 2006 April 12;
      in original form 2006 March 24}

\pubyear{2006}

\begin{document}
\maketitle
\label{firstpage}

\begin{abstract}

Results are presented on component masses and system angular momenta for over a hundred low-temperature contact binaries.
It is found that the secondary components in close binary systems are very similar in mass.
Our observational evidence strongly supports the argument that the 
evolutionary process goes from near-contact binaries to A-type contact binaries, without any need of mass loss from the system. 
Furthermore, the evolutionary direction of A-type into W-type systems with a simultaneous mass and angular momentum loss is also discussed.
The opposite direction of evolution seems to be unlikely, since it requires an increase 
of the total mass and the angular momentum of the system.

\end{abstract}


\section{Introduction}  
\label{sect:intro}
In his two seminal papers, \citet{Lucy1968a,Lucy1968b} 
not only showed that two stars can exist in an envelope of a 
common equipotential and thus resolved the overall 
hitherto unexplainable properties of W~UMa-type stars, but also 
very clearly indicated that such contact binaries must have
very dissimilar components. As pointed out by \citet{Hazle1970}, 
evolution can create this dissimilarity.

The evolutionary state of contact binary stars remains unclear
because their spectra cannot be analyzed for abundances due to 
the extreme broadening and blending of spectral lines. 
Indirect information though, such as their progressively increasing 
numbers with age in old open \citep{Rci1998} and globular 
\citep{Rci2000} clusters, as well as the kinematic characteristics
\citep{GB1988, Bilir2005}, very strongly suggest an advanced age of
$>2$ Gyr.

Recently, \citet{Stepien2004} has developed a model with the
currently less massive component being the more evolved one.
Such a model is conceptually very close to that used to
explain the semi-detached Algols. 
In his model, the current secondary (less-massive)
components must be in some cases very low in mass to explain
systems like AW~UMa or SX~Crv \citep{Rasio1995}.

In this Letter we present a summary of results on the
component masses (for 112 systems) and system angular momenta 
(for 93 systems) for low-temperature contact binaries. 
The sample used was collected mainly from the list of contact binaries 
defined by \citet{Kreiner2003}. Half of the systems have solutions published in the frame of the W~UMa project (papers I-VI)
\citep{Kreiner2003, Baran2004, Zola2004, Gazeas2005, Zola2005, Gazeas2006}.
All the rest were collected from the literature, the physical parameters of which 
have been determined accurately, using photometric 
light curves and radial velocity measurements for both components.
We also included the near-contact binaries (NCBs) and the 
detached close binaries (DCBs) listed in Tables 2 and 3 of \citet{YE2005}, 
in order to compare their physical parameters with those of our sample.

We had to exclude the cases, where the third light and the
low-amplitude light variation give unreliable solutions, 
such as V2150~Cyg, V899~Her, HT~Vir, BL~Eri and GO~Cyg.
In some cases, close binaries in triple systems have led to spurious 
solutions and for this reason they were excluded from our sample.  
Other cases with third light contribution do not seem to 
produce any problem. Such systems, which are members in multiple systems,
have better geometrical configuration and usually have very good spectroscopic determination
of the third light contribution, allowing accurate determination of the orbital and physical
parameters.
Two systems, V351 Peg and V402 Aur were very probably incorrectly classified as 
W-type binaries, although the shape of their light curves does not 
support such a classification. Both systems have equal minima in their light curves, making it 
difficult to distinguish them from each other. Since their physical parameters (masses and periods) were closer to 
those of A-type binaries, they were classified as A-type systems.

Only recently, good spectroscopic data has become
available for more than a hundred contact binaries. Since even small-mass 
secondaries can be observed spectroscopically in
contact systems, most objects have been analyzed
as double-lined binaries bypassing any need of inferences
based on single-lined data or solely on sometimes highly 
unreliable photometric elements (particularly mass ratios). 

In our study we consider component "1" as the more massive one. Our assumption is based on the 
double-lined spectroscopic observations, where the mass ratio is taken as $q=M_{\rmn{2}}/M_{\rmn{1}} <1$.

\section{Masses}
\label{sect:masses}

\begin{figure}
\begin{center}
\rotatebox{0}{\scalebox{0.8}{\includegraphics{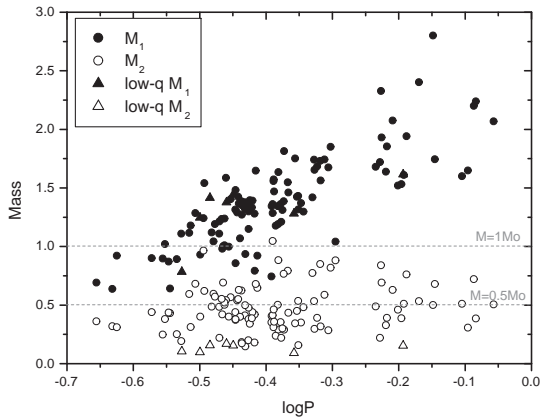}}}
\caption{\label{fig1}
The mass distribution for the two components of all 112 contact binaries in
our sample. The masses of the primary components are plotted with full circles,
while those of secondaries with open circles.
Triangles represent the low-q systems, with secondary components of very small mass.
The absence of systems with periods between the values -0.30 and -0.23 (0.5 and 0.6 days) is obvious.
Note that the masses of the secondaries are between the limits of $0-1M_{\odot}$, 
while the masses of primaries gradually increase, almost proportional to $logP$. 
}
\end{center}
\end{figure}

\begin{figure}
\begin{center}
\rotatebox{0}{\scalebox{0.8}{\includegraphics{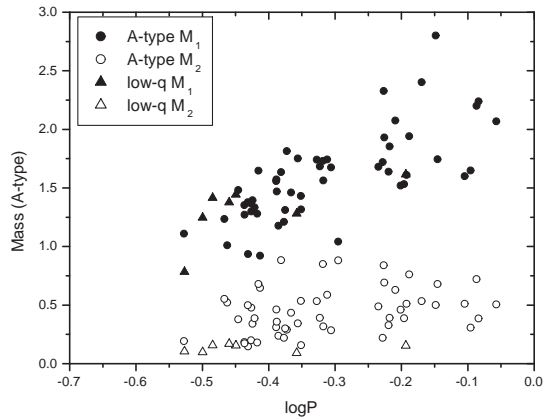}}}
\rotatebox{0}{\scalebox{0.8}{\includegraphics{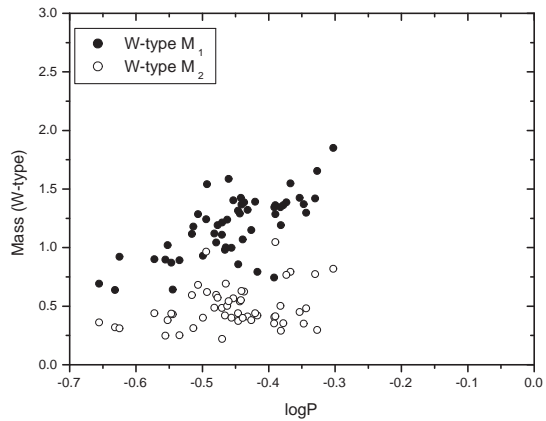}}}
\caption{\label{fig2}
The mass distribution of the two components of 60 A-type contact binaries (upper panel) 
and 52 W-type contact binaries (lower panel) of our sample.
Triangles represent the low-q systems, as in Figure 1.
}
\end{center}
\end{figure}

In Figure~\ref{fig1} we present the distribution of the component masses versus 
the orbital period. It seems that the secondary components in 
all systems are very similar in masses, regardless of the orbital period.
Masses of secondaries are between the values of 0 and 1 $M_{\odot}$.
The mean value of the mass of the secondaries is $0.45M_{\odot}$, while the 
primaries have masses between 0.5 and $2.5M_{\odot}$ 
(only HV~UMa has $M_{\rmn{1}}=2.8M_{\odot}$). 
The same pattern of distribution appears when the masses of 
A and W-type systems are plotted in separate figures (Figures~\ref{fig2}a and ~\ref{fig2}b).
In this case, the average mass of A-type secondaries is equal to 
$0.41M_{\odot}$ and that of W-type equal to $0.49M_{\odot}$.
It is remarkable to see the similarity of the mass distribution of secondary components with 
that of white dwarfs \citep{Madej2004}. According to \citet{Stepien2004}, mass exchange is taking place 
in the majority of contact binaries and the secondaries are helium-rich objects.
In this case the masses of the secondary components are expected to be similar or smaller 
than those of white dwarfs, which could have been grown as cores of isolated stars.

Another interesting feature shown in Figure~\ref{fig1} as well as in the upper and lower panels of Figure~\ref{fig2} 
is that there is a total absence of systems with periods between 0.5 and 0.6 d. 
This gap is rather caused by the selection effect of our sample of contact binaries.
Many contact binaries with equal minima and periods close to 12 h are difficult to be observed and can be mistaken as 
pulsating variables, with a period of 6 h (i.e. $\beta\ Lyrae$ variables), or remain unclassified.
A recent study \citep{Rci2002} has also showed that many contact binaries are still undetected. 

In our sample, all W-type systems have orbital periods shorter than 0.5 d, 
while the A-type systems can have all possible periods in the range considered, with a 
small preference in large values.

Seven systems (CK~Boo, FP~Boo, SX~Crv, GR~Vir, TZ~Boo, AW~UMa and FG~Hya)
(especially AW~UMa, SX~Crv and TZ~Boo) are plotted with triangles in all figures, as 
they are low-q systems with very low-mass secondaries ($M_{\rmn{2}}<0.17M_{\odot}$).
In these systems the rotational angular momentum is mostly "absorbed" from the primary component 
and plays a significant role on the total angular momentum of the system.
According to \citet{Rasio1995} these very low-q systems cannot exist, 
since $J_{\rmn{orb}}>3J_{\rmn{spin}}$.

There is a co-existence of A and W-type systems with 
periods between 0.3 and 0.5 d. All systems with $P < 0.3$ d 
are of W-type and all with $P > 0.6$ d are of A-type.

\begin{figure}
\begin{center}
\rotatebox{0}{\scalebox{0.8}{\includegraphics{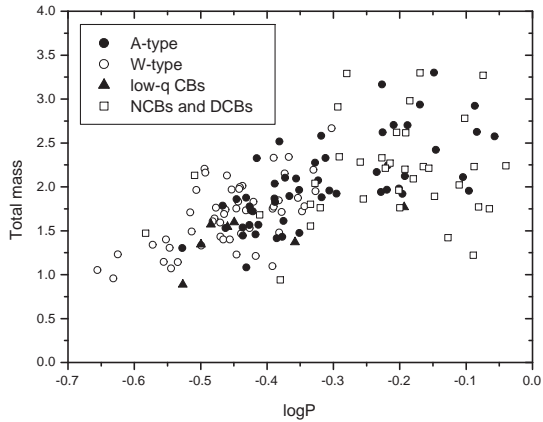}}}
\caption{\label{fig3}
The total mass distribution of 112 contact binaries (full circles for A-type and open circles for W-type), 
25 near contact and 11 detached binaries (open squares). 
Triangles represent the low-q contact binaries, as in Figure 1.
}
\end{center}
\end{figure}

A very interesting feature is shown in Figure~\ref{fig3}, where the 
total mass of the above systems is plotted versus the period. 
It is obvious that A-type systems are in general more massive than W-type ones. 
For comparison, in the same graph, we plotted the total mass of our CBs with the total mass of the
short-period ($<1\ d$) NCBs and DCBs, taken from 
\citet{YE2005}. One can see that most of the NCB and DCB 
systems have total mass similar to that of A-type contact binaries but larger than that of W-type contact binaries.

These observational facts strongly support the evolutionary progress from near-contact into contact configuration
of A-type, without the need of mass loss from the system, as proposed by \citet{YE2005}. Further progress in this direction,
may transform the A-type systems into W-type, with a simultaneous mass loss, 
or can lead the A-type systems to become binaries with extreme small mass ratio.
The opposite direction would require the total mass to increase, which is unreasonable.

\section{Angular momenta}
\label{sect:ang-mom}

\begin{figure}
\begin{center}
\rotatebox{0}{\scalebox{0.8}{\includegraphics{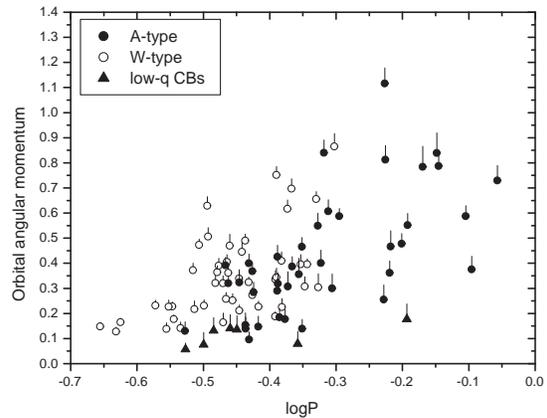}}}
\caption{\label{fig4}
The orbital angular momentum distribution of the 93 contact binaries of our sample (full circles for A-type and open circles for W-type). 
Vertical lines above each point represent the amount of the spin angular momentum
of the two components, which, if added to the orbital, will represent the total angular momentum of each system.
Triangles represent the low-q contact binaries, as in Figure 1.
}
\end{center}
\end{figure}

In Figure~\ref{fig4} we present a plot of the orbital angular momentum versus 
the orbital period. In this plot we see that A-type systems generally tent to have larger 
angular momenta than the W-type systems. 
On the top of each point in the graph, a vertical line is added, representing the
amount of the spin angular momentum of the two components. 
The sum of the spin and orbital angular momenta will give the total angular momentum of each system.

In this way it can be seen that the angular momentum of close binaries
can be studied with either the orbital or the total angular momentum.
The orbital angular momentum is not affected from observational or modeling errors, 
but only from masses and orbital periods. On the other hand, the spin angular momentum
is affected from the errors in mass, radius and from the assumption we make for the radius of gyration,
which is still under investigation \citep{Rasio1995}. 
For example, a slightly smaller radius of gyration would  
shift the seven low-q systems of our sample in a stable state.

Figure~\ref{fig4} shows that formation of W-type systems from A-type is possible to be done, but the opposite direction is not strongly supported. 
Evolution from A-type to W-type systems seems to occure with a simultaneous mass and angular momentum loss, 
unless the evolutionary direction, after the NCBs evolve to contact systems, follows two separate tracks.
However, we cannot exclude the possibility that some A-type systems with small angular momentum 
have evolved from W-type systems, while others (with large angular momentum) have evolved directly from NCBs \citep{YE2005}.

\section{Discussion}
\label{sect:disc}

Do both A and W-type contact binaries have the same origin? 
Is the one type progenitor of the other? The above questions are still 
open for investigation. 

The main result of the present study, taking into account the masses and 
angular momenta of contact binaries, is that the W-type systems cannot produce 
A-type binaries, since angular momentum and mass cannot be 
added to a system, but only lost from it.
It seems more reasonable that A-type systems evolve to W-type systems
by loosing mass and angular momentum. 
A similar evolutionary direction from long to short period binaries \citep{Bilir2005} showed that systems with longer 
periods are kinematically young (age 2 Gyr) in contrast to those with shorter periods (age 8 Gys).

An interesting result extracted from our plots is that the secondary components in close binary systems are very similar in mass.
Scatter of the points in our plots is mainly due to inaccurate photometric and spectroscopic
solutions and/or a possible undetected third light contribution. 
Since the best of the available data is used in our sample, only the third light
could produce a problem and this is why some solutions are excluded from our sample.
A very recent study of \citet{PR2006} about the formation of contact binaries in multiple 
systems, suggests that a large percentage of 
close binaries is formed into triple and multiple systems. In such a case, a small amount of 
unreliability is added to all the solutions, if they are affected from an undetected third light. 

Our observational evidence (Figures 1-4) strongly supports the argument that the 
evolutionary process is from NCBs to A-type contact binaries, without any need of mass loss from the system. 
The next step in this scenario may lead either to a transformation of A-type to W-type systems 
with a simultaneous mass and angular momentum loss, or to A-type systems with extremely low mass ratio.
These systems will eventually merge into a single, fast-rotating object.
The opposite direction of evolution seems to be impossible, since it requires an increase 
of the total mass and angular momentum of the system.

\section{Acknowledgments}
\label{sect:Acknowledgments}

The authors gratefully acknowledge Professor S. Rucinski for his valuable help during the preparation of the manuscript
and for many insightful discussions and new ideas about contact binary 
structure and evolution, as well as the anonymous referee for useful suggestions, which improved the article. 
This project was supported by the Special Account for Research 
Grants No 70/3/7187 (HRAKLEITOS) and 70/3/7382 (PYTHAGORAS) of the
National and Kapodistrian University of Athens, Greece.

\end{document}